\documentclass[twocolumn,superscriptaddress,reprint,nofootinbib,prd]{revtex4-1}%
\usepackage{amssymb}
\usepackage{color}
\usepackage{graphicx}
\usepackage{dcolumn}
\usepackage{bm}
\usepackage[header,title,page,titletoc]{appendix}
\usepackage{amsmath}
\usepackage{bbm}
\usepackage{amsfonts}%
\usepackage[dvipdfm,bookmarks,colorlinks=false]{hyperref}
\providecommand{\U}[1]{\protect \rule{.1in}{.1in}}
\newcommand{\beq}{\begin{eqnarray}}
\newcommand{\eeq}{\end{eqnarray}}
\newcommand{\be}{\begin{equation}}
\newcommand{\ee}{\end{equation}}
\newcommand{\bw}{\begin{widetext}}
\newcommand{\ew}{\end{widetext}}
\newcommand{\ba}{\begin{array}}
\newcommand{\ea}{\end{array}}
\newcommand{\bk}{\mathbf{k}}
\newcommand{\br}{\mathbf{r}}

\newcommand{\bq}{\mathbf{q}}

\begin{document}
\title{Generalized L\"uscher's Formula in Multichannel Baryon-Baryon Scattering}

 \author{Ning Li}
\affiliation{%
School of Physics, Peking University, Beijing 100871, P.~R.~China
}%

 \author{Song-Yuan Li}
\affiliation{%
School of Physics, Peking University, Beijing 100871, P.~R.~China
}%

\author{Chuan Liu}%
\email[Corresponding author. Email: ]{liuchuan@pku.edu.cn}
\affiliation{%
School of Physics and Center for High Energy Physics, Peking
University, Beijing 100871, P.~R.~China
}%

\begin{abstract}
 In this paper,  L\"uscher's formula is generalized to the case of two spin-$\frac{1}{2}$ particles
 in two-channel scattering based on Ref.~\cite{Li:2012bi}. This is first done in
 a non-relativistic quantum mechanics model and then generalized to quantum field theory.
 We show that L\"uscher's formula obtained from these two different methods are equivalent
 up to terms that are exponentially suppressed in the box size. This formalism can
 be readily applied to future lattice QCD calculations.
\end{abstract}


\maketitle

\section{INTRODUCTION}

 The scattering of two baryons is important for the study of strong interaction
 which is one of the four elementary interactions in Nature. The interactions among baryons
 are also relevant for the understanding of nuclear matter which is also crucial in
 other fields of physics. However, due to its non-perturbative nature at low-energies,
  theoretical study of baryon interactions requires non-perturbation methods such as
 lattice chromodynamics (lattice QCD).
 Lattice QCD is a non-perturbative method implemented
 in discretized Euclidean space-time. Within this formalism, physical quantities are
 encoded into various Euclidean correlation functions which in turn can be measured using
 Monte Carlo simulations. Since the lattice calculations are all performed in
 a finite volume, the quantities obtained in lattice simulations need to be
 transformed into physical quantities that are measured in experiments.
 For the study of hadron-hadron interactions at low energies,
 L\"uscher has set up a formalism which relates the
 two-particle elastic scattering $S$-matrix elements with the
 corresponding two-particle energies in a finite volume~\cite{luscher86:finiteb,luscher90:finite,luscher91:finitea,luscher91:finiteb}.
 Since the advent of L\"uscher's formula, various lattice studies, both quenched~\cite{Gupta:1993rn,Fukugita:1994ve,Aoki:1999pt,Aoki:2002in,Liu:2001ss,Du:2004ib,Aoki:2005uf,Aoki:2002ny} and
 unquenched~\cite{Yamazaki:2004qb,Beane:2005rj,PhysRevD.77.094507,PhysRevD.81.074506,Feng2010268,PhysRevD.86.034031,PhysRevD.87.034505,PhysRevD.87.054502,PhysRevLett.111.222001,PhysRevLett.111.192001}, have been performed over the years to investigate the scattering of hadrons.

 The original L\"uscher's formula was derived for two spinless massive particles
 in the center of mass (COM) frame below the inelastic threshold.
 It has been generalized in various ways over the years:
 to the boosted frames~\cite{Rummukainen:1995vs,Kim:2005gf,Gockeler:2012yj}, to the case of asymmetric boxes~\cite{Li:2003jn,Feng:2004ua,Detmold2004170}, with twisted boundary
 conditions~\cite{Bedaque2005208,Doring:2011vk,Briceno:2013hya,Agadjanov:2013wqa},
 to the case of particles with spin, e.g. to the case of two
 baryons~\cite{Ishizuka:2009bx,Detmold2004170,Beane2004106,PhysRevD.76.034502,%
 Beane200555,Bedaque200482,PhysRevD.84.091503,PhysRevD.84.114502}
 and beyond the inelastic threshold~\cite{He:2005ey,Liu:2005kr,Bernard:2010fp,Hansen:2012tf,PhysRevD.88.094507,Guo:2012hv}, and even to three-particle case~\cite{Roca:2012rx,Polejaeva:2012ut,Briceno:2012rv,Hansen:2013dla}.

 In our previous study~\cite{Li:2012bi}, we have generalized
 the formulism to the case of multi-channel scattering of a spinless particle
 with a spin-$1/2$ particle in arbitrary frame.
 In this paper, we  continue to synthesize L\"uscher's formula for the multi-channel two-particle scattering with each of the two hadrons having spin $\frac{1}{2}$.
 We will call them baryons for
 simplicity and they in principle can be of the same type (e.g. two protons)
 or different (e.g. one proton and one neutron).
 Beyond the inelastic threshold, the scattering becomes multi-channel.
 In this paper, we only study the two-channel scattering case.
 It can be generalized readily to other multichannel scattering cases easily.
 In a recent comprehensive study, Briceno has studied the most general two-particle
 scenario~\cite{Briceno:2014oea}.

 The organization of this paper is as follows:
 In Sec. \ref{sec:QM}, we start the discussion in the case of non-relativistic quantum mechanics.
 In Sec. \ref{sec:QF}, generalization to quantum field theory is done. We also compare
 the results obtained within quantum field theory with those obtained in previous section
 and show that they are actually equivalent up to terms that are exponentially suppressed in the box size.
 In Sec. \ref{conclusion}, we will give the conclusions. Possible applications of these
 formulae in future lattice computations are also discussed. Some explicit formulas
 used in this paper are listed in the appendix for reference.

 \section{L\"{u}scher's formula for baryon-baryon scattering in non-relativistic quantum mechanics}
 \label{sec:QM}



 We start our discussion from a non-relativistic quantum mechanics model
 in infinite volume in the continuum.
 The model we use here is the same as that of L\"uscher~\cite{luscher91:finitea} in which a
 finite-ranged potential is assumed. After singling out the center of mass motion,
 one focuses on the relative motion of the two baryons.
 In the center-of-mass (COM) frame, the energy is given by
 \be
 E=\frac{\mathbf{k}_1^2}{2\mu_1} =E_T+\frac{\mathbf{k}_2^2}{2\mu_2}
 \label{energy relation}
 \ee
 where $\mu_1$ and $\mu_2$ is the reduced mass of the
 two-particle system below and above the threshold $E_T$, respectively.
 Since the potential has a finite range,
 in the large $r$ region where the potential vanishes,
 the wave function of the scattering states can be written as
\bw
\beq
\psi_{1;s\nu}(\mathbf{r})\overset{r\rightarrow \infty}{\longrightarrow}\left(
 \ba
 {c}
 \chi_{s\nu}e^{i\bk_1\cdot\br}
 +\sum_{s^{\prime}\nu^{\prime}}\chi_{s^{\prime}\nu^{\prime}}M^{(NR)}_{11;s^{\prime}\nu^{\prime};s\nu}
\frac{e^{ik_1r}}{r}\\
 \sqrt{\frac{\mu_2}{\mu_1}}\sum_{s^{\prime}\nu^{\prime}}\chi_{s^{\prime}\nu^{\prime}}M^{(NR)}_{21;s^{\prime}\nu^{\prime};s\nu}
 \frac{{e^{ik_2r}}}{r}
 \ea
 \right)\;.\label{wave_1}
 \eeq
  \ew
 This wavefunction has the property that, in the infinite past, it reduces to an incident plane wave
 in the first channel.
 The symbol $\chi_{s\nu}$ designates spin-wavefunction which is an eigenstate of spin angular
 momentum of $S^2$ and $s_z$ with the eigenvalues given by $s=0,\nu=0$ (singlet state),
 or $s=1,\nu=1,-1,0$ (triplet state).
 $M^{(NR)}_{if;s^{\prime}\nu^{\prime},s\nu}$ is the scattering amplitude.
 We have added a superscript (NR) to distinguish the
 scattering amplitude introduced here (non-relativistic quantum mechanics) with
 that introduced in quantum field theory later on.
 Subscripts like $i$ and $f$ correspond to the channels and take the value 1 or 2 in this paper.
 There is another analogous but linearly independent wavefunction given by
 \beq
 \psi_{2;s\nu}(\mathbf{r})\overset{r\rightarrow \infty}{\longrightarrow}\left(
 \ba
 {c}
 \sqrt{\frac{\mu_1}{\mu_2}}\sum_{s^{\prime}\nu^{\prime}}\chi_{s^{\prime}\nu^{\prime}}M^{(NR)}_{12;s^{\prime}\nu^{\prime},s\nu}
 \frac{{e^{ik_{1}r}}}{r}\\
 \chi_{s\nu}e^{i\bk_2\cdot\br}
 +\sum_{s^{\prime}\nu^{\prime}} \chi_{s^{\prime}\nu^{\prime}}M^{(NR)}_{22;s^{\prime}\nu^{\prime};s\nu}
\frac{e^{ik_{2}r}}{r}
 \ea \right)\;.\label{wave_2}
 \eeq

 which reduces to an incident plane wave in the second channel.
 Choosing the z-axis to coincide with either $\bk_1$ or $\bk_2$, the scattering amplitudes introduced above are
 related to the $S$-matrix elements as~\cite{Newton}
 \bw
 \beq
 M^{(NR)}_{11;s^{\prime}\nu^{\prime};s\nu}(\mathbf{\hat{k}_{1}\cdot{\hat{r}}})&=&\frac{1}{2ik_{1}}
 \sum_{l^{\prime}=0}^\infty\sum_{l=0}^\infty\sum_{J=l-1}^{l+1}\sqrt{4\pi(2l+1)}i^{(l-l^{\prime})}
 (S_{11;l^{\prime}s^{\prime};ls}^{J}-\delta_{l^{\prime}l}\delta_{s^{\prime}s})\langle{J}M|l^{\prime}m^{\prime};s^{\prime}\nu^{\prime}\rangle
 \langle{J}M|l0;s\nu\rangle{Y_{l^{\prime}m^{\prime}}(\mathbf{\hat{r}})}
 \;,\label{NR_amplitude_1}
 \eeq
 \beq
 M^{(NR)}_{12;s^{\prime}\nu^{\prime};s\nu}(\mathbf{\hat{k}_{2}\cdot{\hat{r}}})
 &=&\frac{1}{2i\sqrt
 {k_{1}k_{2}}}\sum_{l^{\prime}=0}^\infty\sum_{l=0}^\infty\sum_{J=l-1}^{l+1}\sqrt{4\pi(2l+1)}i^{(l-l^{\prime})}
 {S^{J}_{12;l^{\prime}s^{\prime};ls}}\langle{J}M|l^{\prime}m^{\prime};s^{\prime}\nu^{\prime}\rangle
 \langle{J}M|l0;s\nu\rangle{Y_{l^{\prime}m^{\prime}}(\mathbf{\hat{r}})}
 \;. \label{NR_amplitude_2}
 \eeq
 \ew
 Comparing with Eq.~(\ref{wave_1}), Eq.~(\ref{wave_2}), Eq.~(\ref{NR_amplitude_1}) and Eq.~(\ref{NR_amplitude_2}),
 we obtain the following form as:
 \beq
 \label{eq:wavefunction_intermsof_W}
 \psi_{i;s\nu}(\mathbf{r})=\sum_{s^{\prime}JMll^{\prime}}\sqrt{4\pi(2l+1)}W^{J}_{i;l^{\prime}s^{\prime};ls}
 \langle{J}M|l0;s\nu\rangle{Y^{l^{\prime}s^{\prime}}_{JM}(\hat{\mathbf{r}})}\;,
 \eeq
 where $Y^{l^{\prime}s^{\prime}}_{JM}(\hat{\mathbf{r}})$ is the spin spherical harmonics
 whose explicit form is given by
 \beq
 Y^{ls}_{JM}(\hat{\mathbf{r}})&=&\sum_{m\nu}Y_{lm}(\hat{\mathbf{r}})\chi_{s\nu}\langle{J}M|lm;s\nu\rangle.
 \eeq
 In Eq.~(\ref{eq:wavefunction_intermsof_W}), $W^{J}_{i;l^{\prime}s^{\prime};ls}(r)$
 is the radial wave function of the two-particle scattering states.
 In the large $r$ region, they have the following asymptotic forms:
 \bw
 \beq
 W^{J}_{1;l^{\prime}s^{\prime};ls}(r)&=&\left(
 \ba
 [c]{c}
 \frac{1}{2ik_1r}i^{(l-l^{\prime})}[S^{J}_{11;l^{\prime}s^{\prime};ls}e^{ik_1r}+(-1)^{l+1}e^{-ik_1r}\delta_{ll^{\prime}}\delta_{ss^{\prime}}]\\
 \frac{1}{2ir\sqrt{k_1k_2}}\sqrt{\frac{\mu_2}{\mu_1}}i^{(l-l^{\prime})}S^{J}_{21;l^{\prime}s^{\prime};ls}e^{ik_2r}
 \ea
 \right)\;,
 \eeq
 \beq
 W^{J}_{2;l^{\prime}s^{\prime};ls}(r)&=&\left(
 \ba
 [c]{c}
 \frac{1}{2ir\sqrt{k_1k_2}}\sqrt{\frac{\mu_{1}}{\mu_{2}}}i^{(l-l^{\prime})}S^{J}_{12;l^{\prime}s^{\prime};ls}e^{ik_1r}\\
 \frac{1}{2ik_2r}i^{(l-l^{\prime})}[S^{J}_{22;l^{\prime}s^{\prime};ls}e^{ik_2r}+(-1)^{l+1}e^{-ik_2r}\delta_{ll^{\prime}}\delta_{ss^{\prime}}]
 \ea
 \right)\;.
 \eeq
 \ew


 Now we enclose the two-particle system in a cubic box of size $L$ and impose periodic boundary condition.
 In the outer region where the potential vanishes, the wave function becomes
 \be
 \psi(\mathbf{r})=\sum_{s^{\prime}JMll^{\prime}s}
 \left[\sum_{i=1}^{2}F_{i;JMls}W^{J}_{i;l^{\prime}s^{\prime};ls}(r)\right]
 Y_{JM}^{l^{\prime}s^{\prime}}(\mathbf{\hat{r}})\;.\label{wave_3}
 \ee
 On the other hand, also in this region,
 the wave function is a linear superposition of
 the singular periodic solutions, $G_{i;JMls}(\mathbf{r};k_{i}^{2})$,
 of the Helmholtz equation. Thus, we also get
 \beq
 \psi(\mathbf{r})&=&\left(
 \ba
 [c]{c}%
 \overset{1}{\underset{s=0}{\sum}}
 \overset{\infty}{\underset{l=0}{\sum}}\overset{l+1}
 {\underset{J=l-1}{\sum}}\overset{J}{\underset{M=-J}{\sum}}
 \upsilon_{1;JMls}G_{1;JMls}(\mathbf{r};k_{1}^{2})\\
 \overset{1}{\underset{s=0}{\sum}}
 \overset{\infty}{\underset{l=0}{\sum}}\overset{l+\frac{1}{2}}{\underset{J=l-\frac{1}{2}}{\sum}}
 \overset{J}{\underset{M=-J}{\sum}}
 \upsilon_{2;JMls}G_{2;JMls}(\mathbf{r};k_{2}^{2})
 \ea
 \right)\;. \label{wave_4}
 \eeq
 The singular solutions of the Helmholtz equation
 $G_{i;JMls}(\mathbf{r};k_{i}^{2})$ can be expanded in terms
 of spherical harmonics as,
 \bw
 \beq
 G_{i;JMls}&=&\frac{(-1)^lk_i^{l+1}}{4\pi}(Y^{ls}_{JM}n_l(k_ir)+\sum_{J^{\prime}M^{\prime}l^{\prime}}
 \mathcal{M}^{(s)}_{i;JMl;J^{\prime}M^{\prime}l^{\prime}}(k^2_i)
 Y^{l^{\prime}s}_{J^{\prime}M^{\prime}}j_{l^{\prime}}(k_ir))\;,\label{Green}
 \eeq
 \ew
 The explicit form of $\mathcal{M}^{(s)}_{i;JMl;J^{\prime}M^{\prime}l^{\prime}}(k^2_i)$
 can be found, for example, in Ref.~\cite{Li:2012bi}.
 Comparison of Eq.~(\ref{wave_3}) with Eq.~(\ref{wave_4})
 then leads to four linear equations of the coefficients.
 In order to have non-trivial solutions
 for them, the determinant of the corresponding matrix
 must vanish which leads to the basic form of L\"uscher's formula:
 \bw
 \be
 \label{eq:luscher_general_form}
 \left|
 \ba
 {cc}
 \sum_{l^{\prime\prime}}(S^{J}_{11;l^{\prime\prime}s^{\prime};ls}
 -\delta_{ll^{\prime\prime}}\delta_{ss^{\prime}})
 \mathcal{M}^{(s^{\prime})}_{1;JMl^{\prime\prime};J^{\prime}M^{\prime}l^{\prime}}\\
 -i\delta_{JJ^{\prime}}\delta_{MM^{\prime}}(S^{J}_{11;l^{\prime}s^{\prime};ls}+\delta_{ll^{\prime}}\delta_{ss^{\prime}})
 &   \sqrt{\frac{k_2}{k_1}}(\sum_{l^{\prime\prime}}S^{J}_{21;l^{\prime\prime}s^{\prime};ls}
 \mathcal{M}^{(s^{\prime})}_{2;JMl^{\prime\prime};J^{\prime}M^{\prime}l^{\prime}}\\
 &-iS^J_{21;l^{\prime}s^{\prime};ls}\delta_{JJ^{\prime}}\delta_{MM^{\prime}})
 \\
  \sqrt{\frac{k_1}{k_2}}(\sum_{l^{\prime\prime}}S^{J}_{12;l^{\prime\prime}s^{\prime};ls}
  \mathcal{M}^{(s^{\prime})}_{1;JMl^{\prime\prime};J^{\prime}M^{\prime}l^{\prime}}\\
 -iS^{J}_{12;l^{\prime}s^{\prime};ls}\delta_{JJ^{\prime}}\delta_{MM^{\prime}})
 &
\sum_{l^{\prime\prime}}(S^{J}_{22;l^{\prime\prime}s^{\prime};ls}
-\delta_{ll^{\prime\prime}}\delta_{ss^{\prime}})
 \mathcal{M}^{(s^{\prime})}_{2;JMl^{\prime\prime};J^{\prime}M^{\prime}l^{\prime}}\\
 & -i\delta_{JJ^{\prime}}\delta_{MM^{\prime}}(S^{J}_{22;l^{\prime}s^{\prime};ls}+\delta_{ll^{\prime}}\delta_{ss^{\prime}})
 \ea
 \right|
 =0\;.
 \ee
\ew

 In a definite irreducible representation (irrep) of the cubic group,
 the basis vectors are labelled as $|\Gamma,\xi,J,l,s,n\rangle$ where $\Gamma$ denotes the representation;
 $\xi$ runs from 1 to the number of the dimension and $n$ runs from 1 to the multiplicity of
 the representation.
 This basis can be expressed by linear combinations of $|JMls\rangle$
 and the corresponding matrix $\mathcal{M}_i$ is diagonal with respect to $\Gamma$
 and $\xi$ by Schur's lemma~\cite{luscher91:finitea}.
 Therefore, in a particular symmetry sector $\Gamma$, L\"uscher's formula becomes
 \bw
 \be
  \label{eq:luscher_general_form2}
 \left|
 \ba
 {cc}
 \sum_{l^{\prime\prime}}(S^{J}_{11;l^{\prime\prime}s^{\prime};ls}
 -\delta_{ll^{\prime\prime}}\delta_{ss^{\prime}})
 \mathcal{M}^{(s^{\prime})}_{1;Jl^{\prime\prime};J^{\prime}l^{\prime}}(\Gamma)\\
 -i\delta_{JJ^{\prime}}(S^{J}_{11;l^{\prime}s^{\prime};ls}+\delta_{ll^{\prime}}\delta_{ss^{\prime}})
 &   \sqrt{\frac{k_2}{k_1}}(\sum_{l^{\prime\prime}}S^{J}_{21;l^{\prime\prime}s^{\prime};ls}
 \mathcal{M}^{(s^{\prime})}_{2;Jl^{\prime\prime};J^{\prime}l^{\prime}}(\Gamma)\\
 &-iS^J_{21;l^{\prime}s^{\prime};ls}\delta_{JJ^{\prime}})
 \\
  \sqrt{\frac{k_1}{k_2}}(\sum_{l^{\prime\prime}}S^{J}_{12;l^{\prime\prime}s^{\prime};ls}
  \mathcal{M}^{(s^{\prime})}_{1;Jl^{\prime\prime};J^{\prime}l^{\prime}}(\Gamma)\\
 -iS^{J}_{12;l^{\prime}s^{\prime};ls}\delta_{JJ^{\prime}})
 &
\sum_{l^{\prime\prime}}(S^{J}_{22;l^{\prime\prime}s^{\prime};ls}
-\delta_{ll^{\prime\prime}}\delta_{ss^{\prime}})
 \mathcal{M}^{(s^{\prime})}_{2;Jl^{\prime\prime};J^{\prime}l^{\prime}}(\Gamma)\\
 & -i\delta_{JJ^{\prime}}(S^{J}_{22;l^{\prime}s^{\prime};ls}+\delta_{ll^{\prime}}\delta_{ss^{\prime}})
 \ea
 \right|
 =0\;.
 \ee
 \ew
 For the case of integral total angular momentum $J$,
 we need to consider the cubic group $O$
 which has five irrep's: $A_1$, $A_2$, $E$, $T_1$ and $T_2$ with
 dimensions $1$, $1$, $2$, $3$ and $3$.

 For two spin $\frac{1}{2}$ particles, the total spin quantum numbers of the system can
 take $0$ (singlet state) or $1$ (triplet states). The parity of the states depends only
 on orbital angular momentum quantum number and is given by $(-)^l$. Thus, for
 the singlet state and the triplet states with $l=J$, parity is simply $(-)^J$
 while for the other cases it is $(-)^{J+1}$.
 For parity-conserving theories like QCD, there is no scattering between states
 with opposite parity\cite{Newton}.
  Then we should divide the L\"uscher's formula into the case a and case b, they corresponding to the states of parity $(-1)^{J+1}$ and parity $(-1)^{J}$ respectively.

 {\em case a}. $s=s'=1$, $l=J\pm1$\\
  L\"uscher's formula becomes
 \bw
 \be
 \left|
 \ba
 {cc}
 \sum_{l^{\prime\prime}}(S^{J}_{11;l^{\prime\prime}1;l1}
 -\delta_{ll^{\prime\prime}})
 \mathcal{M}^{(1)}_{1;Jl^{\prime\prime};J^{\prime}l^{\prime}}
 -i\delta_{JJ^{\prime}}(S^{J}_{11;l^{\prime}1;l1}+\delta_{ll^{\prime}})
 &   \sqrt{\frac{k_2}{k_1}}(\sum_{l^{\prime\prime}}S^{J}_{21;l^{\prime\prime}1;l1}
 \mathcal{M}^{(1)}_{2;Jl^{\prime\prime};J^{\prime}l^{\prime}}
 -iS^J_{21;l^{\prime}1;l1}\delta_{JJ^{\prime}})
 \\
  \sqrt{\frac{k_1}{k_2}}(\sum_{l^{\prime\prime}}S^{J}_{12;l^{\prime\prime}1;l1}
  \mathcal{M}^{(1)}_{1;Jl^{\prime\prime};J^{\prime}l^{\prime}}
 -iS^{J}_{12;l^{\prime}1;l1}\delta_{JJ^{\prime}})
 &
\sum_{l^{\prime\prime}}(S^{J}_{22;l^{\prime\prime}1;l1}
-\delta_{ll^{\prime\prime}})
 \mathcal{M}^{(1)}_{2;Jl^{\prime\prime};J^{\prime}l^{\prime}}
  -i\delta_{JJ^{\prime}}(S^{J}_{22;l^{\prime}1;l1}+\delta_{ll^{\prime}})
 \ea
 \right|
 =0\;.\label{casea}
 \ee
\ew
 If we consider the explicit parity and assume that the cutoff angular momentum is $\Lambda=4$,
 the the decomposition in this case is as follows:
 $0^-=A^-_1$, $1^+=T_1^+$, $2^-=E^-+T_2^-$, $3^+=A^+_2+T^+_1+T^+_2$, and $4^-=A^-_1+E^-+T^-_1+T^-_2$.
 In the following, we will only focus on $A^-_1$ representation.
 Neglecting the mixing between $J=0$ and $J=4$,
 L\"uscher's formula is given by
 \bw
 \be
 \left|
 \ba
 {cc}
 (S^{0}_{11;11;11}
-1)
 \mathcal{M}^{(1)}_{1;01;01}
  -i(S^{0}_{11;11;11}+1)
 &   \sqrt{\frac{k_2}{k_1}}(S^{0}_{21;11;11}
 \mathcal{M}^{(1)}_{2;01;01}
 -iS^{0}_{21;11;11})
 \\
  \sqrt{\frac{k_1}{k_2}}(S^{0}_{12;11;11}
  \mathcal{M}^{(1)}_{1;01;01}
 -iS^{0}_{12;11;11})
 &
(S^{0}_{22;11;11}
-1)
 \mathcal{M}^{(1)}_{2;01;01}-i(S^{0}_{22;11;11}+1)
 \ea
 \right|
 =0\;.
 \ee
\ew

 {\em case b}: $l=l^{\prime}=J$, $s=0,1$.\\
 For this case, the L\"uscher's formula can be expressed as:
 \bw
 \be
 \left|
 \ba
 {cc}
 \sum_{l^{\prime\prime}}(S^{J}_{11;l^{\prime\prime}s^{\prime};ls}
 -\delta_{ll^{\prime\prime}}\delta_{ss^{\prime}})
 \mathcal{M}^{(s^{\prime})}_{1;Jl^{\prime\prime};J^{\prime}l^{\prime}}\\
 -i\delta_{JJ^{\prime}}(S^{J}_{11;ls^{\prime};ls}+\delta_{ss^{\prime}})
 &   \sqrt{\frac{k_2}{k_1}}(\sum_{l^{\prime\prime}}S^{J}_{21;l^{\prime\prime}s^{\prime};ls}
 \mathcal{M}^{(s^{\prime})}_{2;Jl^{\prime\prime};J^{\prime}l^{\prime}}\\
 &-iS^J_{21;ls^{\prime};ls}\delta_{JJ^{\prime}})
 \\
  \sqrt{\frac{k_1}{k_2}}(\sum_{l^{\prime\prime}}S^{J}_{12;l^{\prime\prime}s^{\prime};ls}
  \mathcal{M}^{(s^{\prime})}_{1;Jl^{\prime\prime};J^{\prime}l^{\prime}}\\
 -iS^{J}_{12;ls^{\prime};ls}\delta_{JJ^{\prime}})
 &
\sum_{l^{\prime\prime}}(S^{J}_{22;l^{\prime\prime}s^{\prime};ls}
-\delta_{ll^{\prime\prime}}\delta_{ss^{\prime}})
 \mathcal{M}^{(s^{\prime})}_{2;Jl^{\prime\prime};J^{\prime}l^{\prime}}\\
 &-i\delta_{JJ^{\prime}}(S^{J}_{22;ls^{\prime};ls}+\delta_{ss^{\prime}})
 \ea
 \right|
 =0\;.\label{Luscher_formula5}
 \ee
\ew
 If we consider the explicit parity and also suppose that the cutoff angular momentum is $\Lambda=4$,
 the decomposition in this case becomes
 $0^+=A^+_1$, $1^-=T_1^-$, $2^+=E^++T_2^+$, $3^-=A^-_2+T^-_1+T^-_2$ and $4^+=A^+_1+E^++T^+_1+T^+_2$.
 Focusing on the $A^{+}_1$ representation which corresponds $J=0$ and
 ignoring the index $l$ and $l^{\prime}$, both of which are unity,
 a similar formula can be easily obtained from Eq.~(\ref{Luscher_formula5}).

 The above discussion has not taken into account the possibility
 for the identical nature of the two particles in which case
 singlet-triplet transition within the same parity is allowed.
 However, if we further assume that the two particles are
 identical, then singlet-triplet transition is forbidden since the singlet state has
 an antisymmetric spin wave function which then requires a symmetric spatial one
 that necessarily has {\em positive} parity while the triplet states have the opposite parity.
 Below, we list L\"uscher formulae for $s=s^{\prime}=0$ and $s=s^{\prime}=1$ cases respectively.
 \bw
 \be
 \left|
 \ba
 {cc}
 \sum_{l^{\prime\prime}}(S^{J}_{11;l^{\prime\prime}0;l0}
 -\delta_{ll^{\prime\prime}})
 \mathcal{M}^{(0)}_{1;Jl^{\prime\prime};J^{\prime}l^{\prime}}\\
 -i\delta_{JJ^{\prime}}(S^{J}_{11;l0;l0}+1)
 &   \sqrt{\frac{k_2}{k_1}}(\sum_{l^{\prime\prime}}S^{J}_{21;l^{\prime\prime}0;l0}
 \mathcal{M}^{(0)}_{2;Jl^{\prime\prime};J^{\prime}l^{\prime}}\\
 &-iS^J_{21;l0;l0}\delta_{JJ^{\prime}})
 \\
  \sqrt{\frac{k_1}{k_2}}(\sum_{l^{\prime\prime}}S^{J}_{12;l^{\prime\prime}0;l0}
  \mathcal{M}^{(0)}_{1;Jl^{\prime\prime};J^{\prime}l^{\prime}}\\
 -iS^{J}_{12;l0;l0}\delta_{JJ^{\prime}})
 &
\sum_{l^{\prime\prime}}(S^{J}_{22;l^{\prime\prime}0;l0}
-\delta_{ll^{\prime\prime}})
 \mathcal{M}^{(0)}_{2;Jl^{\prime\prime};J^{\prime}l^{\prime}}\\
 &-i\delta_{JJ^{\prime}}(S^{J}_{22;l0;l0}+1)
 \ea
 \right|
 =0\;.\label{Lusher_formulae_1}
 \ee

 \be
 \left|
 \ba
 {cc}
 \sum_{l^{\prime\prime}}(S^{J}_{11;l^{\prime\prime}1;l1}
 -\delta_{ll^{\prime\prime}})
 \mathcal{M}^{(1)}_{1;Jl^{\prime\prime};J^{\prime}l^{\prime}}\\
 -i\delta_{JJ^{\prime}}(S^{J}_{11;l1;l1}+1)
 &   \sqrt{\frac{k_2}{k_1}}(\sum_{l^{\prime\prime}}S^{J}_{21;l^{\prime\prime}1;l1}
 \mathcal{M}^{(1)}_{2;Jl^{\prime\prime};J^{\prime}l^{\prime}}\\
 &-iS^J_{21;l1;l1}\delta_{JJ^{\prime}})
 \\
  \sqrt{\frac{k_1}{k_2}}(\sum_{l^{\prime\prime}}S^{J}_{12;l^{\prime\prime}1;l1}
  \mathcal{M}^{(1)}_{1;Jl^{\prime\prime};J^{\prime}l^{\prime}}\\
 -iS^{J}_{12;l1;l1}\delta_{JJ^{\prime}})
 &
\sum_{l^{\prime\prime}}(S^{J}_{22;l^{\prime\prime}1;l1}
-\delta_{ll^{\prime\prime}})
 \mathcal{M}^{(1)}_{2;Jl^{\prime\prime};J^{\prime}l^{\prime}}\\
 &-i\delta_{JJ^{\prime}}(S^{J}_{22;l1;l1}+1)
 \ea
 \right|
 =0\;.\label{Lusher_formulae_2}
 \ee
 \ew
 From these explicit expressions, it is seen that they are quite similar as those in the
 case of meson-meson two-channel scattering~\cite{Liu:2005kr}.

  Finally, let us comment briefly on L\"uscher formula in moving frames (MF).
 In this case, one should expanded the wave function of the system
 in the outer region in terms of modified Green's function $G^{\mathbf{d}}_{i;JMls}(\mathbf{r};k_{i}^{\ast2})$
 and modified matrix $\mathcal{M}^{\mathbf{d}(s)}_{i;JMl;J^{\prime}M^{\prime}l^{\prime}}(k^{\ast2}_i)$~\cite{Rummukainen:1995vs}.
 The explicit forms of these quantities can be obtained by substituting $Y^{l\frac{1}{2}}_{JM}$ for $Y^{ls}_{JM}$,
 and $\langle{J}M|lm;\frac{1}{2}\nu\rangle$ for $\langle{J}M|lm;s\nu\rangle$
 in corresponding formulae in Ref.~\cite{Li:2012bi}.
 L\"uscher's formula takes exactly the same form as Eq.(\ref{casea}) and (\ref{Luscher_formula5})
 except that all the labels of $\mathcal{M}^{(s)}_{i;JMl;J^{\prime}M^{\prime}l^{\prime}}(k^{\ast2}_i)$ are replaced by $\mathcal{M}^{\mathbf{d}(s)}_{i;JMl;J^{\prime}M^{\prime}l^{\prime}}(k^{\ast2}_i)$.
 Apart from the above mentioned substitutions, extra attention should also be paid to the difference in symmetry.
 In order to discuss L\"uscher's formula in MF, we should introduce the space group $\mathcal{G}$,
 which is a semi-direct product of lattice translational group $\mathcal{T}$ and the cubic group $O$.
 The representations are characterized by the little group $\Gamma$
 and the corresponding total momentum $\mathbf{P}$.
 For example, for the cases $\mathbf{P}=\frac{2\pi}{L}\mathbf{e}_3$, $\mathbf{P}=\frac{2\pi}{L}(\mathbf{e}_1+\mathbf{e}_2)$,
 and $\mathbf{P}=\frac{2\pi}{L}(\mathbf{e}_1+\mathbf{e}_2+\mathbf{e}_3)$,
 the corresponding little groups are $C_{4v}$, $C_{2v}$ and $C_{3v}$,
 respectively~\cite{Gockeler:2012yj}.
 Then, following similar steps as in the COM frame, one can easily obtain the explicit
 formulas for these different little groups.

 \section{L\"{u}scher's formula for baryon-baryon scattering in quantum field theory}
 \label{sec:QF}

 In this section, we will discuss the same problem in quantum field theory.
 L\"{u}scher's formula will be obtained in the case of single-channel and then generalized to
 the case of multi-channel, following similar ideas in~\cite{Kim:2005gf,Hansen:2012tf,Li:2012bi}.
 The procedure is similar to that in Refs.~\cite{Kim:2005gf,Hansen:2012tf,Li:2012bi} so we
 will be quite brief here and the reader is directed to those references for further details.

 It is well-known that two-particle spectrum of the system can be determined
 from appropriate correlation functions:
 \be
 \label{eq:CP_definition}
 C(P)=\int_{L;x}e^{i(-\mathbf{P}\cdot \mathbf{x}+Ex^{0})}\langle0|\sigma(x)\sigma^\dagger(0)|0\rangle\;,
 \ee
 where $P=(E,\mathbf{P})$ is the total
 four-momentum of the two-particle system. The interpolating operator
 $\sigma(x)$ is chosen to have an overlap with the
 two-particle states in question and $\int_{L;x}=\int_{L}d^{4}x$ stands for
 the space-time integration over the finite volume.
 Two-particle spectra correspond to poles of $C(P)$ in the $E$ plane.
 The correlation function $C(P)$ may be expressed in terms of
 Bethe-Salpeter amplitude $iK$~\cite{Kim:2005gf,Hansen:2012tf,Li:2012bi}:
 \beq
 \label{eq:sigma_BS}
 & &C(P)=\int_{L;q}\sigma_{q}\left[Z_{1}\Delta_{1}\otimes{Z_{2}}\Delta_{2}\right]_{q}
 \sigma_{q}^\dagger\nonumber\\
 &+&
 \int_{L;q,q^{\prime}}\sigma_{q}\left[Z_{1}\Delta_{1}\otimes{Z}_{2}\Delta_{2}\right]_{q}
 iK_{q,q^{\prime}}\left[Z_{1}\Delta_{1}\otimes{Z}_{2}\Delta_{2}\right]_{q^{\prime}}\sigma_{q^{\prime}}^{\dagger}+\cdots\nonumber\\
 \eeq
 where we have adopted the notation for the two-particle propagators:
 \beq
 Z_n(q)\Delta_n(q)&=&\int{d^4xe^{iqx}\langle\psi_n(x)\bar{\psi}_n(0)\rangle}
 \eeq
 with $n=1,2$ denoting two particles whose propagators are given by
 \beq
 \Delta_n(q)=\frac{i(q^{\mu}\gamma_{\mu}+m_n)}{q^2-m_n^2+i\epsilon}
 \eeq
 where $m_1$ and $m_2$ are the mass of the two baryons.
 The notation $Z_{1}\Delta_{1}\otimes{Z_{2}}\Delta_{2}$ in Eq.~(\ref{eq:sigma_BS})
 denotes a direct product in Dirac space since each of the propagators is
 a spinor in this space.
 Following similar steps as in Refs.~\cite{Kim:2005gf,Hansen:2012tf,Li:2012bi},
 the correlation function can be separated into
 two parts: $C(P)=C^\infty(P)+C^{FV}(P)$, with the $C^\infty(P)$
 being the infinite-volume limit contribution while the second part $C^{FV}(P)$
 being the finite-volume corrections that contains the finite-volume two-particle poles,
 which is what we are interested in.
 The explicit expression of $C^{FV}$ may be written as
 \beq
 C^{FV}&=&-A^{\prime}FA+A^{\prime}F(iM)FA+\cdots
 \nonumber\\
 &=&-A^{\prime}F\frac{1}{1+iMF}A \label{finite_corralator_1}
 \eeq
 where $A$/$A'$ is the Bethe-Salpeter amplitude for the initial/final state,
 $F$ represents the factor associated with the two-particle loop integration/summation
 and $M$ is the scattering amplitude.
 In the COM frame, we denote
 \be
 \label{eq:def_of_FMtilde}
 F=C(\bq^{\ast})\tilde{F}\;,\;\;\;
 M=C(\bq^{\ast})^{-1}\tilde{M}\;.
 \ee
 where the definition of $C(\bq^{\ast})$ can be found in
 the appendix c.f. Eq.~(\ref{C_defination}). Basically, the interpolating operator $\sigma_{q}$ is the product of
 two interpolating operators for the two hadrons and without loss of generality,
 we could assume each of the two particles has definite
 parity. Then, we insert parity projection operator $\mathcal{P}=\mathcal{P}_{1\pm}\otimes{\mathcal{P}_{2\pm}}$
 with $\mathcal{P}_{1\pm}=\frac{1\pm\gamma_0}{2}$ and $\mathcal{P}_{2\pm}=\frac{1\pm\gamma_0}{2}$ in
 the correlation function (\ref{eq:sigma_BS}).
 Thus, $C(\mathbf{q^{\ast}})$ can be viewed as a $4\times4$ matrix,
 and Eq.~(\ref{finite_corralator_1}) becomes
  \begin{align}
 C^{FV}  &  =-AFA^{\prime}+AF(iM)FA^{\prime}+\ldots\nonumber\\
 &  =-AC(\mathbf{q^{\ast}})\tilde{F}A^{\prime}+AC(\mathbf{q^{\ast}})\tilde{F}(i\tilde{M})\tilde{F}A^{\prime}+\ldots\nonumber\\
 &  =-AC(\mathbf{q^{\ast}})\tilde{F}\frac{1}{1+i\tilde{M}\tilde{F}}A^{\prime} \;.
 \label{finite_corralator_2}
 \end{align}
  L\"uscher's formula can be obtained by requiring the finite
 correlator in Eq.~(\ref{finite_corralator_2}) to have divergent
 eigenvalues. Thus the so-called quantization condition is
 \beq
 \label{eq:general_formula}
 \det(1+i\tilde{M}\tilde{F})=0\;.
 \eeq
 In order to compare the formulae obtained in quantum field
 theory and those in non-relativistic quantum mechanics,
 we also need the relation between the two versions of the scattering
 matrix $\tilde{M}$ and $M^{NR}$.
 This has been obtained in Ref.~\cite{Li:2012bi} which we quote here:
 \beq
 \tilde{M}_{JMls;J^{\prime}M^{\prime}l^{\prime}s^{\prime}}&=&
 8\pi{E^{\ast}}M^{(NR)}_{JMls;J^{\prime}M^{\prime}l^{\prime}s^{\prime}}
 \;.
 \label{field_scattering_amplitude}
 \eeq
 The quantity $M^{NR}_{JMls;J^{\prime}M^{\prime}l^{\prime}s^{\prime}}$ is
 further related to the $S$-matrix elements via
 \beq
 M^{(NR)}_{JMls;J^{\prime}M^{\prime}l^{\prime}s^{\prime}}
 &=&\delta_{JJ^{\prime}}\delta_{MM^{\prime}}\frac{1}{2ik}i^{l-l^{\prime}}
 (S^{J}_{l^{\prime}s^{\prime};ls}-\delta_{ll^{\prime}}\delta_{ss^{\prime}})\;.
 \label{eq:M_JLM}
 \eeq

 As we have discussed, the two baryon system can form a singlet or a triplet state
 in terms of spin. For parity-conserving theories, these operators are categorized into two cases:

 {\em case a}. $s=s^{\prime}=1$, $l=J\pm1$.

 In this case, we can expand the amplitude $A$/$A'$,
 loop factor $\tilde{F}$ and the scattering amplitude $\tilde{M}$ in
 terms of spin spherical functions:
 \be
 \label{eq:A_s_expand1}
 \left\{ \begin{aligned}
 A({\mathbf{\hat{k}}^{\ast}})&=\sqrt{4\pi}A_{JMl1}Y_{JM}^{l1\dagger}(\mathbf{\hat{k}}^{\ast})
 \\
 A^{\prime}(\mathbf{\hat{k}^{\ast}})&=\sqrt{4\pi}A^{\prime}_{JMl1}Y_{JM}^{l1}(\mathbf{\hat{k}}^{\ast})
 \end{aligned} \right.\;,
 \ee
 and
 \be
 \label{eq:FM_expand1}
 \left\{ \begin{aligned}
 \tilde{F}(\mathbf{\hat{k}}^{\ast},\mathbf{\hat{k}^{\ast \prime}})&=\frac{-1}{4\pi}\tilde{F}_{JMl1;J^{\prime}M^{\prime}l^{\prime}1}
 Y_{JM}^{l1}
 (\mathbf{\hat{k}^{ \ast}})
 Y_{J^{\prime}M^{\prime}}^{l^{\prime}1{\dagger}}(\mathbf{\hat{k}^{\ast \prime}})
 \\
 \tilde{M}(\mathbf{\hat{k}}^{\ast},\mathbf{\hat{k}^{\ast \prime}})&={4\pi}\tilde{M}_{JMl1;J^{\prime}M^{\prime}l^{\prime}1}
 Y_{JM}^{l1}
 (\mathbf{\hat{k}^{\ast}})Y_{J^{\prime}M^{\prime}}^{l^{\prime}1{\dagger}}
 (\mathbf{\hat{k}^{\ast \prime}})
 \end{aligned} \right.
 \ee
 Note that, in the above expressions,
 $Y_{JM}^{l1}(\mathbf{\hat{k}}^{\ast})$ only
 has two components with $s=1$ and  $\nu=\pm 1$.
 When substituting these expansions into the general formula~(\ref{eq:general_formula}),
 one obtains
 \bw
 \beq
 \det(\sum_{l^{\prime\prime}}i^{(l^{\prime}-l^{\prime\prime})}(S^{J}_{l^{\prime\prime}1;l1}
 -\delta_{ll^{\prime\prime}})
 F^{FV(1)}_{1;Jl^{\prime\prime};J^{\prime}l^{\prime}}
 -i\delta_{JJ^{\prime}}(S^{J}_{l^{\prime}1;l1}+\delta_{ll^{\prime}}))&=&0
 \eeq
 \ew

 So far, the discussion has been in the case of single channel scattering.
 Generalization to the multi-channel case is straightforward~\cite{Li:2012bi,Hansen:2012tf}.
 Take the two-channel case as an example,
 the amplitudes $A$ and $A'$ in Eq.~(\ref{eq:A_s_expand1}) become two-component vectors
 while both $\tilde{F}$ and $\tilde{M}$ in Eq.~(\ref{eq:FM_expand1}) become $2\times2$ matrices
 in the so-called ``channel space". L\"uscher's formula then becomes
 \bw
 \be
 \left|
 \ba
 {cc}
 \sum_{l^{\prime\prime}}i^{(l^{\prime}-l^{\prime\prime})}(S^{J}_{11;l^{\prime\prime}1;l1}
 -\delta_{ll^{\prime\prime}})
 F^{FV(1)}_{1;Jl^{\prime\prime};J^{\prime}l^{\prime}}\\
 -i\delta_{JJ^{\prime}}(S^{J}_{11;l^{\prime}1;l1}+\delta_{ll^{\prime}})
 &   \sqrt{\frac{k_2}{k_1}}(\sum_{l^{\prime\prime}}i^{(l^{\prime}-l^{\prime\prime})}S^{J}_{21;l^{\prime\prime}1;l1}
 F^{FV(1)}_{2;Jl^{\prime\prime};J^{\prime}l^{\prime}}
 -iS^J_{21;l^{\prime}1;l1}\delta_{JJ^{\prime}})
 \\\\
  \sqrt{\frac{k_1}{k_2}}(\sum_{l^{\prime\prime}}i^{(l^{\prime}-l^{\prime\prime})}S^{J}_{12;l^{\prime\prime}1;l1}
 F^{FV(1)}_{1;Jl^{\prime\prime};J^{\prime}l^{\prime}}
 -iS^{J}_{12;l^{\prime}1;l1}\delta_{JJ^{\prime}})
 &
\sum_{l^{\prime\prime}}i^{(l^{\prime}-l^{\prime\prime})}(S^{J}_{22;l^{\prime\prime}1;l1}
-\delta_{ll^{\prime\prime}})
 F^{FV(1)}_{2;Jl^{\prime\prime};J^{\prime}l^{\prime}}
 \\&-i\delta_{JJ^{\prime}}(S^{J}_{22;l^{\prime}1;l1}+\delta_{ll^{\prime}})
 \ea
 \right|
 =0\;.
 \ee
\ew

{\em case b}. $l=l^{\prime}=J$, $s=0,1$.

 In this case, we could also expand the amplitudes, loop factor
 and the scattering amplitude in terms of spin spherical functions.
 The single-channel case formula read:
 \be
 \label{eq:A_s_expand2}
 \left\{ \begin{aligned}
 A({\mathbf{\hat{k}}^{\ast}})&=\sqrt{4\pi}A_{JMls}Y_{JM}^{ls\dagger}(\mathbf{\hat{k}}^{\ast})
 \\
 A^{\prime}(\mathbf{\hat{k}^{\ast}})&=\sqrt{4\pi}A^{\prime}_{JMls}Y_{JM}^{ls}(\mathbf{\hat{k}}^{\ast})
 \end{aligned} \right.\;,
 \ee
 and
 \be
  \label{eq:FM_expand2}
   \left\{ \begin{aligned}
 \tilde{F}(\mathbf{\hat{k}}^{\ast},\mathbf{\hat{k}^{\ast \prime}})&=\frac{-1}{4\pi}\tilde{F}_{JMls;J^{\prime}M^{\prime}l^{\prime}s^{\prime}}
 Y_{JM}^{ls}
 (\mathbf{\hat{k}^{ \ast}})
 Y_{J^{\prime}M^{\prime}}^{l^{\prime}s^{\prime}{\dagger}}(\mathbf{\hat{k}^{\ast \prime}})
 \\
 \tilde{M}(\mathbf{\hat{k}}^{\ast},\mathbf{\hat{k}^{\ast \prime}})&={4\pi}\tilde{M}_{JMls;J^{\prime}M^{\prime}l^{\prime}s^{\prime}}
 Y_{JM}^{ls}
 (\mathbf{\hat{k}^{\ast}})Y_{J^{\prime}M^{\prime}}^{l^{\prime}s^{\prime}{\dagger}}(\mathbf{\hat{k}^{\ast \prime}})
  \end{aligned} \right.
 \ee
 Here $Y_{JM}^{ls}(\mathbf{\hat{k}}^{\ast})$ only have two components with $s=1, \nu=0$, and $s=0, \nu=0$,
 and $J=l, J^{\prime}=l^{\prime}$
 When substituting these explicit expressions of $\tilde{M}$ and $\tilde{F}$ into
 the general formula~(\ref{eq:general_formula})
 and follow similar steps as in {\em case a},
 we can obtain the two-channel scattering L\"uscher's formula as
\bw
 \be
 \left|
 \ba
 {cc}
 \sum_{l^{\prime\prime}}i^{(l^{\prime}-l^{\prime\prime})}(S^{J}_{11;l^{\prime\prime}s^{\prime};ls}
 -\delta_{ll^{\prime\prime}}\delta_{ss^{\prime}})
 F^{FV(s^{\prime})}_{1;Jl^{\prime\prime};J^{\prime}l^{\prime}}\\
 -i\delta_{JJ^{\prime}}(S^{J}_{11;ls^{\prime};ls}+\delta_{ss^{\prime}})
 &   \sqrt{\frac{k_2}{k_1}}(\sum_{l^{\prime\prime}}i^{(l^{\prime}-l^{\prime\prime})}S^{J}_{21;l^{\prime\prime}s^{\prime};ls}
 F^{FV(s^{\prime})}_{2;Jl^{\prime\prime};J^{\prime}l^{\prime}}\\
 &-iS^J_{21;ls^{\prime};ls}\delta_{JJ^{\prime}})
 \\
  \sqrt{\frac{k_1}{k_2}}(\sum_{l^{\prime\prime}}i^{(l^{\prime}-l^{\prime\prime})}S^{J}_{12;l^{\prime\prime}s^{\prime};ls}
  F^{FV(s^{\prime})}_{1;Jl^{\prime\prime};J^{\prime}l^{\prime}}\\
 -iS^{J}_{12;ls^{\prime};ls}\delta_{JJ^{\prime}})
 &
\sum_{l^{\prime\prime}}i^{(l^{\prime}-l^{\prime\prime})}(S^{J}_{22;l^{\prime\prime}s^{\prime};ls}
-\delta_{ll^{\prime\prime}}\delta_{ss^{\prime}})
 F^{FV(s^{\prime})}_{2;Jl^{\prime\prime};J^{\prime}l^{\prime}}\\
 &-i\delta_{JJ^{\prime}}(S^{J}_{22;ls^{\prime};ls}+\delta_{ss^{\prime}})
 \ea
 \right|
 =0\;
 \ee
 \ew
 In this case, if the two particles are identical, there is no transition between the $s=0, \nu=0$
 and $s=1, \nu=0$,
 So the above L\"uscher's formulae reduce to their counterparts
 in the case of the meson-meson multichannel scattering~\cite{Hansen:2012tf}.

 We are now in a position to compare the formulae obtained using quantum field theory here
 and those in non-relativistic quantum mechanics in the previous section.
 Recall that (see the comments at the
 end of previous section), in the latter case, L\"uscher's formula is expressed
 in terms of the function
 $\mathcal{M}^{\mathbf{d}(s)}_{i;JMl;J^{\prime}M^{\prime}l^{\prime}}(k^{\ast2}_i)$
 while in the quantum field theory case it is expressed in terms of
 $F^{FV(s)}_{i;JMl;J^{\prime}M^{\prime}l^{\prime}}$.
 It turns out that these two quantities are related by:
 \be
 \label{eq:relation_F_and_M}
 F^{FV(s)}_{i;JMl;J^{\prime}M^{\prime}l^{\prime}}=
i^{l}\mathcal{M}^{\mathbf{d}(s)}_{i;JMl;J^{\prime}M^{\prime}l^{\prime}}
i^{-l^{\prime}}
 \;,
 \ee
 as discussed in the appendix.
 This relation is valid up to terms that are suppressed exponentially
 by the box size. With this relation in mind, it becomes clear then that
 these two versions of L\"uscher's formula
 are equivalent up to terms that are exponentially
 suppressed in the box size.
 As mentioned in the introduction, Briceno recently performed the most
 general study of two-particle system with arbitrary spin in a finite volume~\cite{Briceno:2014oea}.
 We have compared our results with his 
 and agreements are found when comparable.

\section{conclusions}
\label{conclusion}

 In this paper, based on our previous works~\cite{Li:2012bi}, we continue
 to discuss multi-channel L\"uscher's formula for two-particle system with spin $\frac{1}{2}$.
 We have done this in the cases of both non-relativistic quantum mechanics and  quantum field theory.
 Finally, we show that the two versions of L\"uscher's formula
 obtained within two different methods are equivalent up
 to terms that are exponentially suppressed in the box size.
 Our formula can be readily utilized in the study of baryon-baryon
 scattering, especially in the case of multi-channel scattering.
 A typical example would be $N\Sigma-N\Lambda$ coupled channel scattering
 which is important for the study of dense nuclear matter~\cite{PhysRevLett.109.172001}.

  \section*{Acknowledgments}

 The authors would like to thank Dr. R.~Briceno for his critical comments to
 the early versions of this manuscript. This work is supported in part by the
 National Science Foundation of China (NSFC) under the project
 No.11335001. It is also supported in part by the DFG and the NSFC (No.11261130311) through funds
 provided to the Sino-Germen CRC 110 ``Symmetries and the Emergence
 of Structure in QCD''.

\appendix
\section{Loop integration for the single channel}

 This appendix serves to fill the gap that leads to Eq.~(\ref{eq:relation_F_and_M})
 which is utilized in the text to show the equivalence of L\"uscher's formula
 obtained in non-relativistic quantum mechanics model and in quantum field theory.
 To setup the relation~(\ref{eq:relation_F_and_M}), we need to analyze
 intermediate loop integration/summation that appears in the Bethe-Salpeter equation.
 We have adopted the same notation as in Ref.~\cite{Li:2012bi} and
 the reader is directed to that reference for unexplained notations.
 The derivation here is also quite brief and the details closely follows
 those in Ref.~\cite{Li:2012bi}.

 First, we quote a useful formula for the summation from Refs.~\cite{Kim:2005gf,Li:2012bi}.
 Below, quantities with a $*$ indicate the
 corresponding values in the center-of-mass (COM) frame.
 The summation formula we need is:
\beq
S(\mathbf{q}^{\ast})&\equiv & {1\over L^3}\sum_\bk {w^*_\bk\over w_\bk}{f^*(\bk)\over q^2-\bk^{*2}}
\\
&=&\mathcal{P}\int\frac{d^3k^{\ast}}{(2\pi)^3}\frac{f^{\ast}(\mathbf{k}^{\ast})}{q^{\ast2}-k^{\ast2}}
+\sum_{l,m}f_{lm}C^{P}_{lm}(q^{\ast2})
 \label{eq:summation_formula_1}
\eeq
where
\beq
  \label{eq:summation_formula_2}
f^{\ast}(\mathbf{k}^{\ast})&=&\sum^{\infty}_{l=0}\sum^{l}_{m=-l}f^{\ast}_{lm}(k^{\ast})k^{\ast{l}}\sqrt{4\pi}Y_{lm}(\hat{\mathbf{k}}^{\ast})
\eeq
and
\beq
 \label{eq:summation_formula_3}
C^{P}_{lm}(q^{\ast2})&=&\frac{1}{L^3}\sum_{\mathbf{k}}\frac{w^{\ast}_{\mathbf{k}}}{w_{\mathbf{k}}}
\frac{e^{\alpha(q^{\ast2}-k^{\ast2})}}{q^{\ast2}-k^{\ast2}}k^{\ast{l}}\sqrt{4\pi}Y_{lm}(\hat{\mathbf{k}}^{\ast})
\nonumber\\
& &-\mathcal{P}\int\frac{e^{\alpha(q^{\ast2}-k^{\ast2})}}{q^{\ast2}-k^{\ast2}}
\eeq
where $\mathcal{P}$ stands for the principal-value prescription.
 The above formulae will be utilized when we transform the intermediate loop
 integral/summation in the following.

 Following similar steps as in Ref.~\cite{Li:2012bi},
 we need to calculate a loop integration/summation
 in the correlation function as
 \bw
\beq
\label{eq:I_explicit}
I&=&\frac{-1}{L^3}\sum_{\bk}\int\frac{dk_0}{2\pi}
\frac{f(k_0,\bk)(k^{\mu}\gamma_{\mu}+m_1)\otimes((P^{\nu}-k^{\nu})\gamma_{\nu}+m_2)}
{(k^2-m_1^2+i\epsilon)((P-k)^2-m_2^2+i\epsilon)}\nonumber\\
\eeq
 \ew
 In the process of this summation, we have to introduce
 a cutoff function $f(k_0,\bk)$ whose ultraviolet behavior renders
 the expressions convergent. The explicit form of this function
 is irrelevant but we will demand that $f(k_0,\bk)$ is an even function of $\bk$,
 i.e. $f(k_0,-\bk)=f(k_0,\bk)$.
 We can then divide the loop integration into two parts. The part that contains the two-particle
 poles in the finite volume is of interest here.
 Performing the $k^0$ integration the expression~(\ref{eq:I_explicit})
 will pick up the relevant poles in the energy plane.
 Following similar steps as in Ref.~\cite{Li:2012bi},
 we obtain the function $C(\bk)$ that appeared in Eq.~(\ref{eq:def_of_FMtilde}),
 which in the COM frame becomes
 \be
 \label{C_defination}
 C(\bk^{\ast})=[m_1I\otimes{E^{\ast}}\gamma_{0}-(k^{\ast\mu}\gamma_{\mu})\otimes
 (k^{\ast\nu}\gamma_{\nu})+m_1m_2]|_{k^{0\ast}=w_{1\mathbf{k}^{\ast}}}
 \ee
 with $w_{1\bk}=\sqrt{\bk^2+m^2_1}$.
 Using the summation formulae~(\ref{eq:summation_formula_1}),(\ref{eq:summation_formula_2}), (\ref{eq:summation_formula_3}) quoted at
 the beginning of this appendix, we find that
 the finite volume correction part of the loop integral~(\ref{eq:I_explicit}), $I_{FV}$,
 is given by
 \beq
 I_{FV}&=&\frac{q^{\ast}f_{00}^{\ast}(q^{\ast})C(\mathbf{q^{\ast}})}{8\pi{E^{\ast}}}
 -\frac{iC(\mathbf{q^{\ast}})}{2E^{\ast}}\sum_{lm}f_{lm}^{\ast}(q^{\ast})C_{lm}^{P}(q^{\ast2})\;.\nonumber\\    \label{I_fv}
 \eeq
 Now we proceed as in Ref.~\cite{Li:2012bi}. By using the completeness of spherical harmonics,
 we are able to setup the following relations:
 \bw
 \beq
 \label{field_F3}
 F&\equiv &C(\mathbf{q^{\ast}})\tilde{F}
 =\frac{q^{\ast}C(\mathbf{q^{\ast}})}{8\pi{E^{\ast}}}(1+iF^{FV})\;,
 \eeq
  \beq
 \tilde{F}_{JMls;J^{\prime}M^{\prime}l^{\prime}s^{\prime}}&=&\frac{q^{\ast}}{8\pi
 {E^{\ast}}}(\delta_{JJ^{\prime}}\delta_{MM^{\prime}}\delta_{ll^{\prime}}\delta_{ss^{\prime}}
 +iF_{JMls;J^{\prime}M^{\prime}l^{\prime}s^{\prime}}^{FV})\;,    \label{field_F2}
 \eeq
 \beq
 F_{JMls;J^{\prime}M^{\prime}l^{\prime}s^{\prime}}^{FV}&=&\frac{-4\pi}{q^{\ast}}
 \sum_{l_1m_1}\frac{\sqrt{4\pi}}{q^{\ast l_{1}}}{C_{l_1m_1}^{P}}\int{d\Omega^{\ast}}Y_{JM}^{ls\dagger}
 Y_{l_1m_1}^{\ast}Y_{J^{\prime}M^{\prime}}^{l^{\prime}s^{\prime}}\nonumber\\
 &=&F_{JMl;J^{\prime}M^{\prime}l^{\prime}}^{FV(s)}\delta_{ss^{\prime}};, \label{field_F1}
 \eeq
 \beq
 C^P_{l_1m_1}(q^{{\ast}2})&=&\frac{1}{L^3}\sum_{\mathbf{k}}\frac{w_{1\mathbf{k}}^{\ast}}{w_{1\mathbf{k}}}
 \frac{e^{\alpha(q^{{\ast}2}-k^{{\ast}2})}}{q^{{\ast}2}-k^{{\ast}2}}k^{{\ast}l_1}\sqrt{4\pi}Y_{l_1m_1}
 -\mathcal{P}\int\frac{d^3k^{\ast}}{(2\pi)^3}\frac{e^{\alpha(q^{{\ast}2}-k^{{\ast}2})}}{q^{{\ast}2}-k^{{\ast}2}}\;,
 \eeq
 \ew

 In Refs.~\cite{Briceno:2012yi,Li:2012bi},
 for the scattering of two particles,
 the relations between the quantity $C^P_{lm}(q^{\ast2})$
 and the function $Z^{\mathbf{d}}_{lm}(1,\mathbf{\kappa}^2)$ were established.
 For the two channel case, this relation reads
\beq
C^P_{i;lm}(q_i^{\ast2})&=&-\frac{\sqrt{4\pi}}{\gamma{L^3}}(\frac{2\pi}{L})^{l-2}Z^{\mathbf{d}}_{lm}(1,\mathbf{\kappa}^2_i)
\eeq
 Using this relation, we then arrive at the following equality:
\beq
F^{FV(s)}_{i;JMl;J^{\prime}M^{\prime}l^{\prime}}&=&
i^{l}\mathcal{M}^{\mathbf{d}(s)}_{i;JMl;J^{\prime}M^{\prime}l^{\prime}}
i^{-l^{\prime}}\label{proved_2}
\eeq
 which is exactly Eq.~(\ref{eq:relation_F_and_M}) utilized in the text.
These formulas hold up to terms that are vanishing exponentially in the box size.

\bibliographystyle{apsrev4-1}
\bibliography{reference}

\end{document}